\documentclass[11pt]{article}
\usepackage{lineno}
\usepackage{times}
\usepackage{textcomp}
\usepackage{amsmath}
\usepackage{amssymb}
\usepackage{bm}
\usepackage{color}
\usepackage{soul}
\usepackage{graphicx}
\usepackage{epstopdf}
\DeclareGraphicsExtensions{.png}
\newenvironment{sciabstract}{%
\begin{quote} \bf}
{\end{quote}}

\newcounter{lastnote}

\title{Complete characterisation of the azimuthal and radial indices of light fields carrying orbital angular momentum}

\author
{Michael Mazilu,$^{1\ast}$ Areti Mourka,$^{1}$ Tom Vettenburg,$^{1}$ Kishan Dholakia$^{1,2}$\\
\\
\normalsize{$^1$ SUPA, School of Physics and Astronomy, University of St Andrews, }\\
\normalsize{St Andrews, Fife, KY16 9SS, UK}\\
\normalsize{$^{2}$College of Optical Sciences, The University of Arizona}\\
\normalsize{1630 East University Boulevard, Tucson, AZ 85721-0094, USA}\\
\\
\normalsize{$^\ast$To whom correspondence should be addressed; E-mail:  michael.mazilu@st-andrews.ac.uk}
}

\newcommand{\micm}{\mbox{$\,$\textmu m}} 

\begin{document} 




\maketitle 
\begin{sciabstract}
The direct determination of the complete transversal state of an electromagnetic field and  accompanying mode indices is essential for the proper quantification of all light-matter interactions~\cite{Alexandrescu:2006p2439}. In particular light fields with cylindrical symmetry such as Laguerre-Gaussian beams can possess orbital angular momentum~\cite{Allen:1992p10107}, central to a wide range of emergent applications in quantum cryptography~\cite{MolinaTerriza:2001p9973}, manipulation~\cite{Grier:2003p7266}, astrophysics~\cite{Berkhout:2008p8343}, microscopy~\cite{hell_nat} and electron beams~\cite{Verbeeck:2010is}.   A wide array of diffractive structures such as arrays of pinholes~\cite{Berkhout:2008p8343}, triangular apertures~\cite{Mourka:2011iy,Hickmann:2010ju}, slits~\cite{Ferreira:2011is,Sztul:2006p8127}, and holograms~\cite{Berkhout:2010um,Berkhout:2011bv} have all recently been used to measure the azimuthal index $\ell$ of individual LG beams. However, all these approaches measure only one single degree of freedom of LG beams, neglecting the radial component or $p$ index and are thus not applicable for a priori unknown beams. Furthermore, it is unclear  which is the optimal aperture and scheme needed to determine simultaneously the azimuthal and radial indices nor the extent with which such an aperture can tolerate deviations in beam parameters. Here, we demonstrate a powerful approach to simultaneously measure the radial and azimuthal indices ($\ell$ and $p$) of both pure and mixed (superposed) LG light fields. We show that the shape of the diffracting element used to measure the mode indices is in fact of little importance and the crucial step is ``training'' any diffracting optical system and transforming the observed pattern into uncorrelated variables (principal components). Modest fluctuations in beam parameters such as waist size and alignment variations can be tolerated in our scheme.  Importantly, beam superpositions can be experimentally decomposed delivering the intensity of each mode and their relative phases.  Our results demonstrate the first complete characterisation of LG beams. The approach can be expanded to other families of beams such as Bessel or Hermite-Gaussian beams and represents a powerful method for characterising the optical multi-dimensional Hilbert space~\cite{MolinaTerriza:2001p9973}. 
\end{sciabstract}

The last two decades have seen immense interest in light fields possessing orbital angular momentum. In particular the orthonormal basis set of Laguerre-Gaussian (LG) beams has been of interest in this regard (see supplement S1). An LG beam is typically characterised by two indices, namely the azimuthal index $\ell$  which denotes the number of cycles of $2\pi$ phase change around the mode circumference and the radial index $p$ where $p$+1 denotes the number of rings present within the light field (counting the central lobe). The azimuthal index leads to an orbital angular momentum of $\ell\hbar$ per photon. 
 A suite of methods have emerged in the last few years that attempt to measure the azimuthal index of an LG light field many based on the far field pattern generated by a diffracting aperture. However, it is to be noted, common to all of these schemes is the fact that the incident light field is assumed to possess a radial index $p=0$. In contrast, in virtually all experimental realisations of LG beams, we are presented with light fields that are appropriately described as superpositions of LG modes, each of the same index $\ell$ but of different $p$ index~\cite{Ohtake:2007ha}. 
Thus it is crucial to include the influence of both mode indices on the diffraction pattern of any aperture or slit and further ascertain whether using any diffracting aperture we are able to determine both the azimuthal and radial indices simultaneously.  Furthermore, open questions include the optimal aperture to use and red the robustness of  the determination of mode indices in the presence of any mis-alignment.  The generic approach we present here addresses all of these issues. 

Our approach is to consider the intensity profile of an LG beam after diffraction from a mask or filter. This intensity profile is in general complicated and all the methods presented here are composed of a training or calibration step in which the response of the optical  diffracting system is measured for every single LG beam considered. The second step corresponds to the actual identification or measurement of an unknown LG beam delivering simultaneously its radial and azimuthal indices. The LG beams are created using a spatial light modulator (SLM) that is wavefront corrected using the optical eigenmode (OEi) technique (see Supplement S2 and S3).


To illustrate our generic approach, we  begin with a discussion of the far-field diffraction pattern of a general LG beam from an aperture composed of three concentric triangular slits. This is an extension of the single triangular aperture already considered, which is presently one of the more powerful techniques for determining the azimuthal index~\cite{Mourka:2011iy,Hickmann:2010ju} by simply counting the number of lobes in the far-field diffraction image. Superposing multiple such apertures might appear, at first glance, a method to determine both the radial and azimuthal indices of an incident field. Indeed, the choice of three concentric apertures aims to probe the radial index of LG beams with a commensurable beam waist. From figure \ref{fig:f1}b (and model presented in supplement S1) we deduce that the diffraction pattern from the triple triangular slit aperture does not offer any simple way to determine the $\ell$ and $p$ beam parameters. However, we remark that regardless of the radial index, the pattern orientation does depend on the sign of the azimuthal index. Although, it cannot be excluded that there exists a specifically designed mask that would  deliver a simple rule for the detection of both $\ell$ and $p$, this is not the case for the triple triangular slit aperture. 

Importantly, the deduction of a complicated $\ell$ and $p$ retrieval rule can be replaced considering face-recognition algorithm readily employed in biometric identification. Here, we choose to use the principal component analysis (PCA) approach to determine the largest variations between the different far-field diffraction patterns from our aperture. The first step of the procedure corresponds to creating a database of all the possible beams that need to be detected. After subtraction of the common mean intensity, we calculate the covariance matrix of these intensity patterns. Its eigenvector with the largest eigenvalue is termed the first ``eigenface'' corresponding to the largest variability of the LG beam training set. In the same way, one can introduce the second ``eigenface'' as the image corresponding to the second largest eigenvalue of the covariance matrix. As can be seen from figure~\ref{fig:f1}d, projecting the measured beams onto these ``eigenfaces'' delivers the first and subsequent principal component representation of the measure. We remark that in this representation the LG beams having the same $\ell$ and $p$ form very tight clusters. Finally, representing the diffraction pattern of an unknown beam in the same way, one can use a classification algorithm to determine the membership of beams with unknown mode indices. For simplicity, we chose the nearest neighbour measure to classify~\cite{Cover:1967jq} but other methods such as the Mahalanobis distance~\cite{Draper:2003p4667} may also be used. Figure~\ref{fig:f1}f shows the classification results displaying a 100\% efficiency i.e. all unknown beams have been correctly identified. 

It is straight forward to deduce the radial index of an LG beam from its intensity profile by simply counting the number of rings in the beam profile. Unfortunately, the complete characterisation of LG beams is more difficult when it has a non zero azimuthal index and thus possesses a topological vortex. Theoretically, the intensity profile of a beam with $\ell$ azimuthal index is identical to the $-\ell$ case. This makes it impossible to deduce the  azimuthal index from the intensity of the beam. The triangular aperture breaks this symmetry and any mask that is not inversion symmetric would be able to distinguish between the two different signs of $\ell$. More generally, $\it{any}$ random mask or aperture can be used to break this symmetry and its diffraction pattern can be used to detect simultaneously both beam indices. Figure~\ref{fig:f2}a shows that the diffraction patterns do not present any prominent features at all and that any distinct LG beam results in a different  diffraction pattern of similar overall form. Consequently, the variations, in the plane of the first two principal components, are more evenly distributed  not privileging any specific radial index as is the case for the triple triangular slit aperture. This is due to the uniformity of the random mask over which the beam extends for different radial indices. Furthermore, it can be noted that the use of the random mask presents an experimental benefit as all diffraction patterns have similar peak intensity enabling the intensity acquisition with similar exposure durations. Using the random mask we achieve, as indeed for the the triple triangular slit aperture, 100\% classification efficiency while eliminating the need to match beam waist and radial index to the size of the triangles used. 


The fundamental question of the interdependence between the radial index, beam alignment and the beam waist is also interesting.  Importantly, the azimuthal index has only a significance in relationship with the beam axis position and direction while the radial index with the beam waist parameter. We therefore investigate the influence of the beam waist fluctuation and beam mis-alignment on the classification ability of our scheme.  Using the PCA detection method, outlined above, it is possible to study the effect of the variation of these parameters by controllably changing the amplitude profile of the LG beam generating SLM~\cite{Lavery:2011tr}. Figure~\ref{fig:f3} shows the results when considering the effect of these parameter fluctuations. The first effect is a clear widening of the scattering cluster of each given beam parameter. This is understandable as beam parameter fluctuations naturally induce a certain variability of the intensity profile. Nevertheless, the correct detection can still be achieved provided that either a larger training set is considered and/or a larger dimensionality of the data is allowed by taking more principal components into account. This last point is illustrated in figure~\ref{fig:f3}e where the detection efficiency, defined by the trace of the confusion matrix normalised to the total number of unknown beams considered, is evaluated as a function of the number principal components.  

\begin{figure}[htbp] 
   \centering
   \includegraphics[scale=.35]{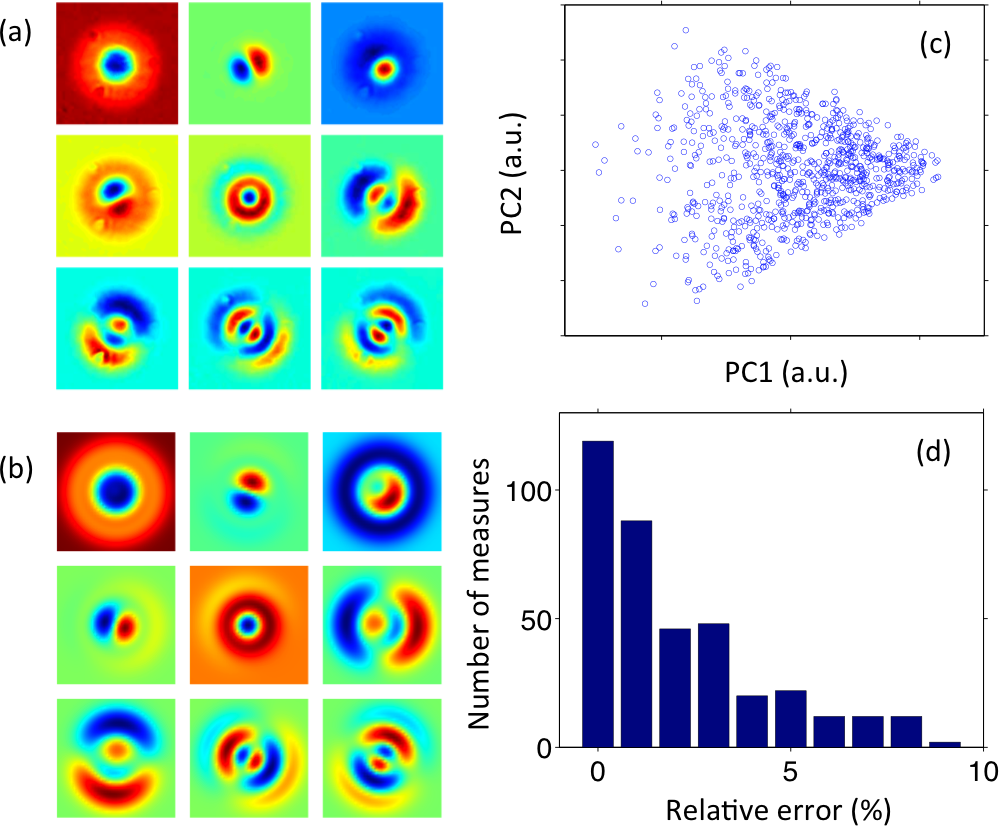} 
   \caption{First nine (a) experimental and (b) theoretical principal components deduced from 900 random  LG beam superposition ($\ell=[0,1]$ and $p=[0,1]$).   (c) First two principle component projections. (d) Individual beam detection error for a different set of 100 random superpositions. The intensity error is defined as $(c_i-m_i)^2$ where $c_i$ and $m_i$ are the, encoded and respectively measured, real amplitude superposition coefficients. }
   \label{fig:f4}
\end{figure}

In quantum optics, entanglement  requires an analysis of superpositions of states. Such superpositions may also occur in the classical domain, for example when considering light fields with fractional azimuthal index~\cite{Gotte:2008p8180}. It is therefore interesting to  extend our approach to the more general case of a superposition of multiple LG beams having different relative amplitudes and phases~\cite{Vaziri:2002p3308}. To that end, we trained the PCA eigenface algorithm on a set of 900 random complex superpositions (keeping the total intensity constant) of four LG beams. A startling observation is that 97\% of the whole training-set's variability can be accounted for by  only the first nine principal components (fig.~\ref{fig:f4}a,b).
Indeed, the superposition is described by four complex numbers leading to four real amplitudes and three relative phases. The non-linear relationship between these degrees of freedom and the far-field diffraction pattern leads to the nine linear degrees of freedom observed experimentally (fig.~\ref{fig:f4}a) and numerically (fig.~\ref{fig:f4}b). 
Further, the projections onto these first principal components forms a dense cloud showing a non-normal distribution due to the constant intensity constraint (fig.~\ref{fig:f4}c). Finally, we can identify the amplitude distribution of an unknown beam by using nearest neighbour classification. Experimentally, the best results were achieved when averaging over the six closest neighbours (see error histogram fig.~\ref{fig:f4}d). The relative phase between the constituent beams can be retrieved by using an optical eigenmode~\cite{DeLuca:2011jl,Mazilu:2011uf} GS algorithm~\cite{GS}  (for more details see supplement S4). 

In this letter, we have presented a straight forward approach  to determine both the radial and azimuthal indices of LG beams by measuring their far-field diffraction pattern from random matrices. This method is robust and can even tolerate a certain degree of beam mis-alignment and beam waist variations. Additionally, our method is able to measure the amplitudes of the different LG components in a superposition of LG beams. Whilst we concentrate here on the LG family of optical beams, the approach we present is generic and can be readily applied to the detection of other family of beams such as Hermite-Gaussian or Bessel beams. Finally, with suitable training, one could envisage the use of this approach to detect low order aberrations that can be described by the family of Zernike polynomials. The mode determination method may be applied beyond the field of electromagnetic waves to sound and matter waves, for example, electron beams. In future work, we will expand the method to the single photon case, enabling the encoding, decoding and manipulation of the radial and azimuthal degrees of freedom of single quanta of light.  

\subsubsection*{Methods}
{\footnotesize  A Helium Neon laser source ($\lambda$ = 633 nm, $P_{max}$ = 5 mW) was used. The laser beam was additionally sent through a 50 \micm\ pinhole in order to obtain a beam featuring a homogeneous Gaussian intensity profile. The laser beam was then subsequently expanded with a telescope ($L_1$ and $L_2$) in order to slightly overfill the chip of a spatial light modulator (SLM) which was a Holoeye LC-R 2500. The SLM operated in the standard first-order configuration and was used to imprint the vortex phase on the incident beam where the LG beam was created in the far-field of the SLM. In order to filter the first order beam, carrying the vortex, from the unmodulated zero-order beam, a pinhole aperture was located in the back focal plane of lens $L_3$. Lens $L_4$ was then used to image the $\ell$ \& $p$ mode beam on the triple triangular slit aperture, with an aperture's size roughly matching the beam waist. A static triple triangular slit aperture imprinted onto photographic film was used and measurements were performed also using a random transparent medium aperture. 
Finally, lens $L_5$ served to create the far-field diffraction pattern in the respective back focal plane where a CCD camera (Basler pi640-210gm, pixel size: 7.4 \micm\ $\times$ 7.4 \micm) was used, in order to record and save the images of the far-field diffraction patterns onto the hard drive of a computer (see supplement for experimental setup figure). To record distinct images, an adjustment between both the size and the thickness of the aperture to the size of the beam waist of the LG beam should be done. Since the diameter of the bright ring depends on the azimuthal index, $\ell$, and the number of nodal rings to the radial index, $p$, respectively. The diffraction patterns are blurred and deformed if the size of the aperture is not commensurate with the diameter of the bright ring. Moreover, the pattern intensity is larger for a large thickness while the pattern gets more distinct for a smaller thickness~\cite{Yongxin:2011iu}. We have adjusted the CCD camera exposure time in order to best highlight the pattern morphology for each recorded pattern.  }

\bibliographystyle{ieeetr}
\bibliography{../papers}

\subsubsection*{Acknowledgements}
We acknowledge helpful critical discussions with Prof. E. M. Wright. We thank the UK Engineering and Physical Sciences Research Council for funding. Kishan Dholakia is a Royal Society-Wolfson Merit Award Holder.

\subsubsection*{Author Contributions}
K.D. and M.M. developed and planned the project. M. M. performed the bulk of the analysis, along with optical modelling and algorithm design. A. M. and T. V.  performed the experimental work.  All authors contributed to the writing of the manuscript.

\newpage

\begin{figure}[p]
   \centering
  \includegraphics[scale=.3]{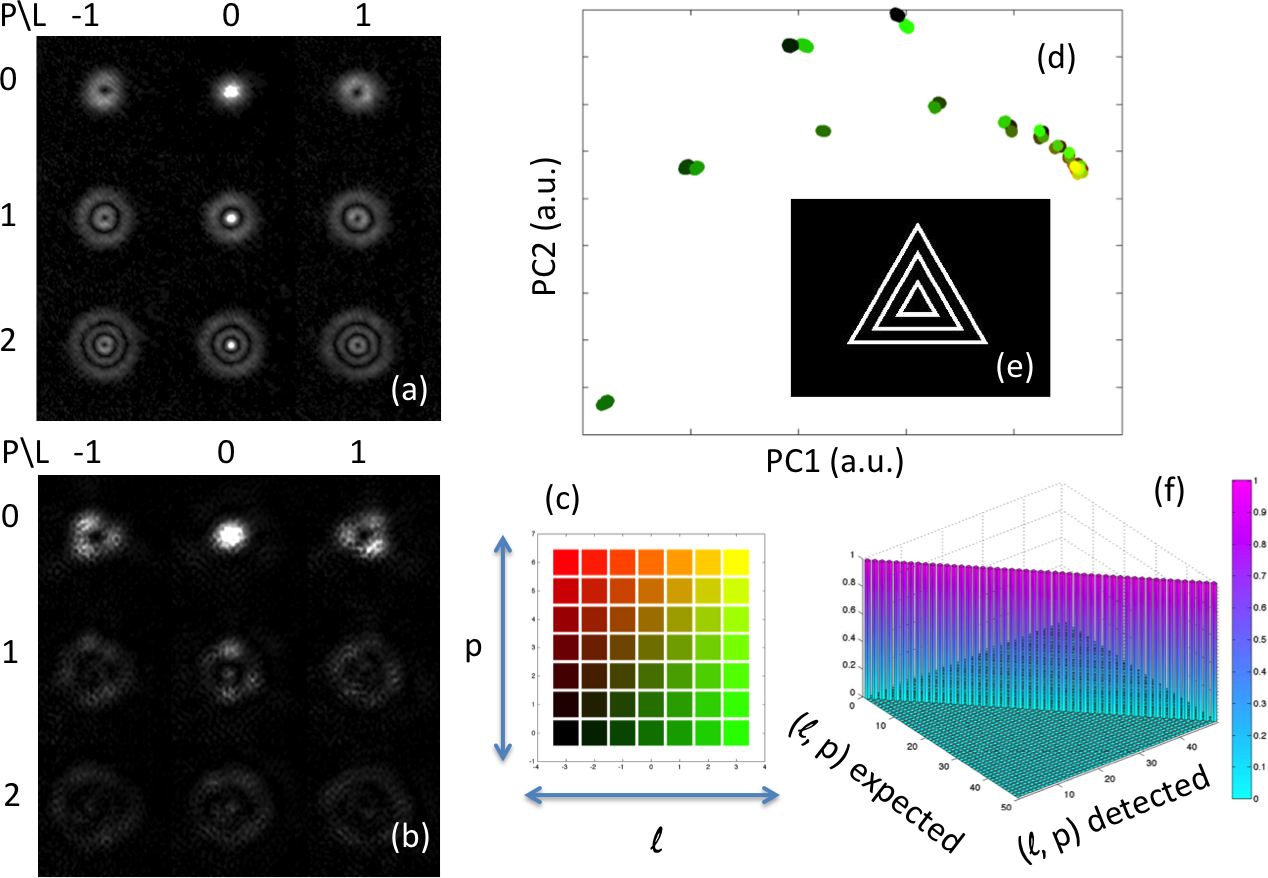} 
   \caption{(a) Table showing the 9 lowest LG modes ($\ell=[-1,1]$ and $p=[0,2]$) and their (b) far field diffraction pattern of the (e) triple triangular slit aperture. (d) Principle components of the diffraction patterns (c) colour coded for the azimuthal and radial LG index. (f) 3D bar chart of the confusion matrix linking expected and detected beam indices.}
   \label{fig:f1}
\end{figure}

\begin{figure}[htbp] 
   \centering
   \includegraphics[scale=.3]{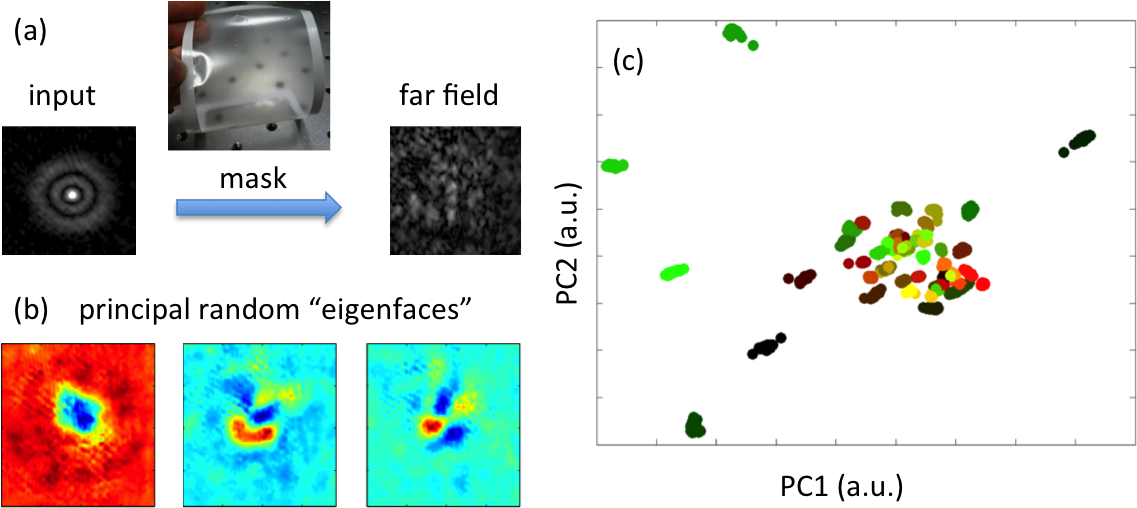} 
   \caption{(a) Experimental implementation of the random mask. (b) First three principal  ``eigenfaces'' based on the far field diffraction patterns ($\ell=[-3,3]$ and $p=[0,6]$) from the random mask. (c) Principal components using the same colour coding as in figure~\ref{fig:f1}.}
   \label{fig:f2}
\end{figure}

\begin{figure}[htbp] 
   \centering
   \includegraphics[scale=.3]{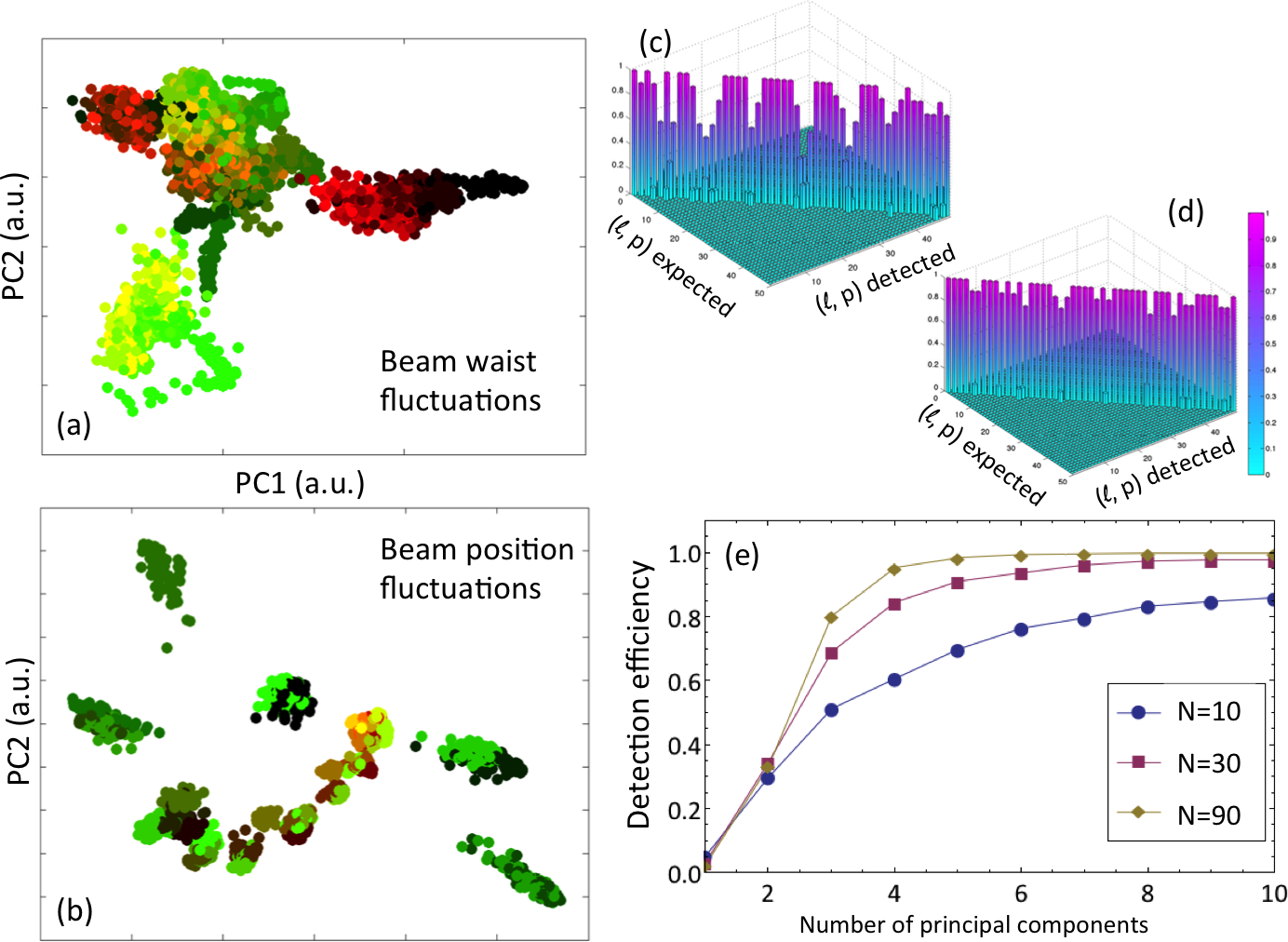} 
   \caption{(a,b) Principle components using the same colour coding as in figure~\ref{fig:f1} for (a) the beam waist fluctuations of 10\% relative to waist and (b) beam position fluctuations of 10\% relative to the waist size. (c,d) Respective confusion matrices for the same cases. (e) Detection efficiency as a function of the number of principal components used for the beam parameter identification where N is the number of far field diffraction images used as a training set for the PCA. }
   \label{fig:f3}
\end{figure}

\clearpage\newpage



\section*{Supplementary information}

\subsection*{S1 Far field diffraction from triangular apertures of Laguerre-Gaussian beams} 

In the focal plane (defined as $z=0$), Laguerre-Gaussian beams are described by the scalar field
\begin{eqnarray}
u^\ell_p(x,y,z=0)&=&(-1)^p\sqrt{\frac{2}{\pi w^2_0}}\sqrt{\frac{p!}{(|\ell |+p)!}} \left(\frac{\sqrt{2} (x\pm iy)}{w_0}\right)^{|\ell |} \nonumber \\
&&\exp\left(-\frac{x^2+y^2}{w_0^2}\right)L_p^{|\ell |} \left(\frac{2(x^2+y^2)}{w_0^2}\right) \label{eq:LGB}
\end{eqnarray}
where $x$, $y$ and $z$ are Cartesian coordinates with $z$ the direction of propagation, $w_0$ corresponds to the Gaussian beam waist, $p$ to the radial index and $\pm$ to the sign of the azimuthal index $\ell$. Further, $L_p^{|\ell|}$ are the generalised Laguerre polynomials. These beams are characterised by the total phase change, equal to $2 \pi \ell$, acquired when integrating the local phase change on a closed loop around the optical axis ($x=y=0$). Additionally, the radial index determines the number of radial nodes to be $p+1$ determined by the roots of the Laguerre polynomial.

The Laguerre-Gaussian beams, as defined by (\ref{eq:LGB}), form a complete orthonormal basis over the transverse $xy$-plane
\begin{equation}
\int u^{\ell'}_{p'}(x,y)u^{*\ell}_p(x,y) dxdy = \delta_{\ell\ell'}\delta_{pp'}
\label{eq:ortho}
\end{equation}
where $\delta_{nn'}$ is the Kronecker delta function. Any transversal beam profile can therefore be decomposed into a sum of Laguerre-Gaussian beams. In general, this sum includes terms having nonzero azimuthal and radial index. Only very specific beam symmetries and profiles allow for a decomposition into beams that involve only  Laguerre-Gaussian modes with $p=0$. This is the main motivation into the study of the detection of general Laguerre-Gaussian beams using the triple triangular slit aperture. 

In general, a paraxial optical system is defined by the ABCD matrix describing the changes in beam profile and phase front curvature as the beam propagates through the system. 
This ABCD matrix defines the propagation through the generalized Huygens-Fresnel integral of the form:
\begin{eqnarray}
u_{2}(x_{2},y_2)&=&\frac{ik}{2\pi B}\int \exp\left(-i\frac{k}{2B}(Ax_{1}^{2}-2x_{1}x_{2}+Dx_{2}^{2})\right) \nonumber \\
&&\exp\left(-i\frac{k}{2B}(Ay_1^2-2y_{1}y_{2}+Dy_{2}^{2})\right)u_{1}(x_{1},y_1) dx_{1}dy_{1} 
\label{eq:huy}
\end{eqnarray}
where $u_{1}(x_{1},y_1)$ and $u_{2}(x_{2}),y_2$ are respectively the fields
 in the initial and final transverse planes. The coefficients A, B, C and
D are obtained through the matrix product of all associated matrices
of the optical elements involved.

After the spatial light modulator, we propagate through a single lens to form the far field
in its focal plane. The associated ABCD matrix is given by
\begin{equation}
\left(\begin{array}{cc}
A & B\\
C & D\end{array}\right)
=
\left(\begin{array}{cc}
1 & f\\
0 & 1\end{array}\right)\left(\begin{array}{cc}
1 & 0\\
-1/f & 1\end{array}\right)\left(\begin{array}{cc}
1 & f\\
0 & 1\end{array}\right)
=
\left(\begin{array}{cc}
0 & f\\
-1/f & 0\end{array}\right)
\label{eq:ABCD}\end{equation}
Here, we remark that the Huygens-Fresnel integral (\ref{eq:huy}) simplifies to a 2D Fourier transform for this specific ABCD matrix.
Finally, the correct diffraction pattern is achieved by setting $u_{1}(x_{1},y_1)=T(x_1,y_1) u^\ell_p(x,y)$ where $T(x_1,y_1)$ corresponds to the transmission function which is only $100\%$ transmissive  in the domains that the triangular slit is transparent. 

\begin{figure}[htbp] 
   \centering
   \includegraphics[scale=.3]{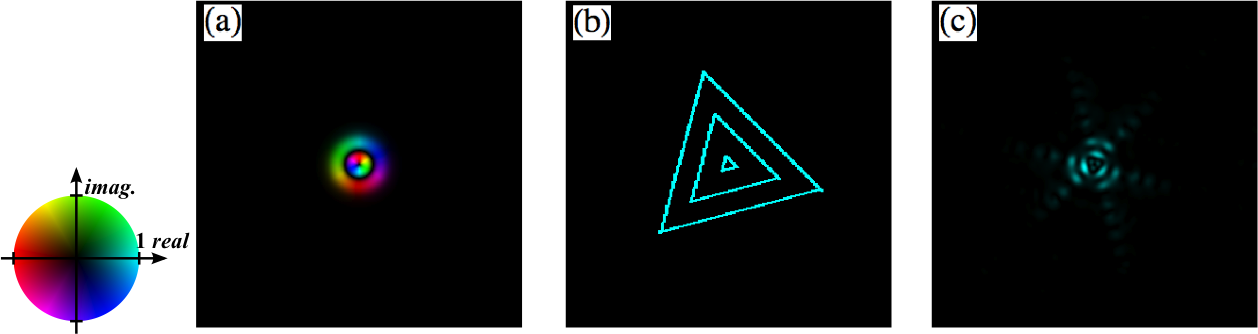} 
   \caption{(a) Cross section of a Laguerre-Gaussian beam ($p=1$, $\ell=1$). The field amplitude is represented using the brightness and the phase using the hue of the colour. (b) Triple triangular slit aperture. (c) Far-field intensity diffraction pattern of the Laguerre-Gaussian beam represented in part (a) diffracting of the triple triangular slit aperture represented in part (b).}
   \label{fig:L1P1}
\end{figure}

For the numerical implementation of the far-field diffraction pattern, we used the above mentioned Fourier transform property. The scaler field of each Laguerre-Gaussien beam is discretised on a square grid and apertured by the triple triangular slit mask. The choice of the mask is based on the ability of the single triangular aperture to distinguish between different azimuthal indices. Additionally, we need to probe with the same mask the number of radial nodes present in the general Laguerre-Gaussian profile. Figure~\ref{fig:L1P1} shows  an example Laguerre-Gaussian beam, the triple triangular slit and the associated far field diffraction pattern. 

\begin{figure}[htbp] 
   \centering
  \includegraphics[scale=.3]{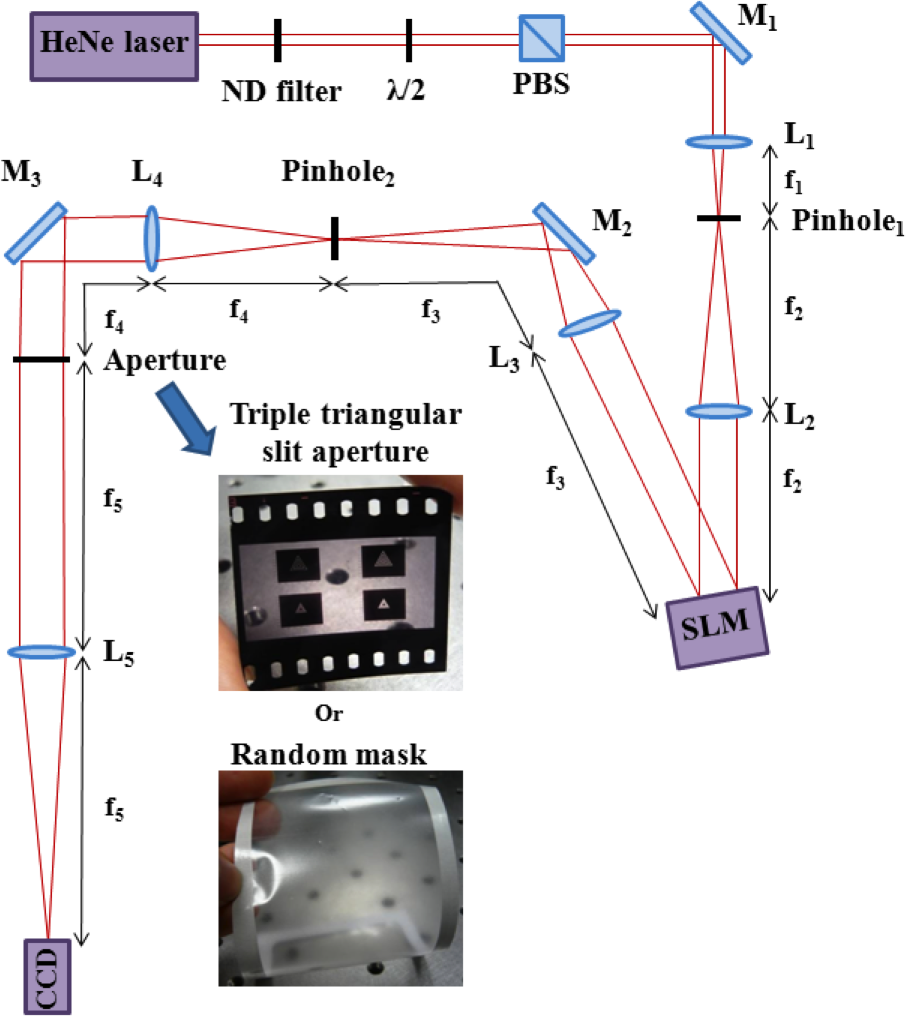} 
   \caption{Schematic of the experimental set-up: L = lens, SLM = spatial light modulator, CCD = charge coupled device camera, PBS = polarizing beam splitter. Focal widths of lenses: $f_1$ = 25mm, $f_2$ = 100cm, $f_3$ = 680mm, $f_4$ = 400mm and $f_5$ = 800mm. A static triple triangular slit aperture was used. }
   \label{fig:setup}
\end{figure}

\begin{figure}[htbp] 
   \centering
   \includegraphics[scale=.35]{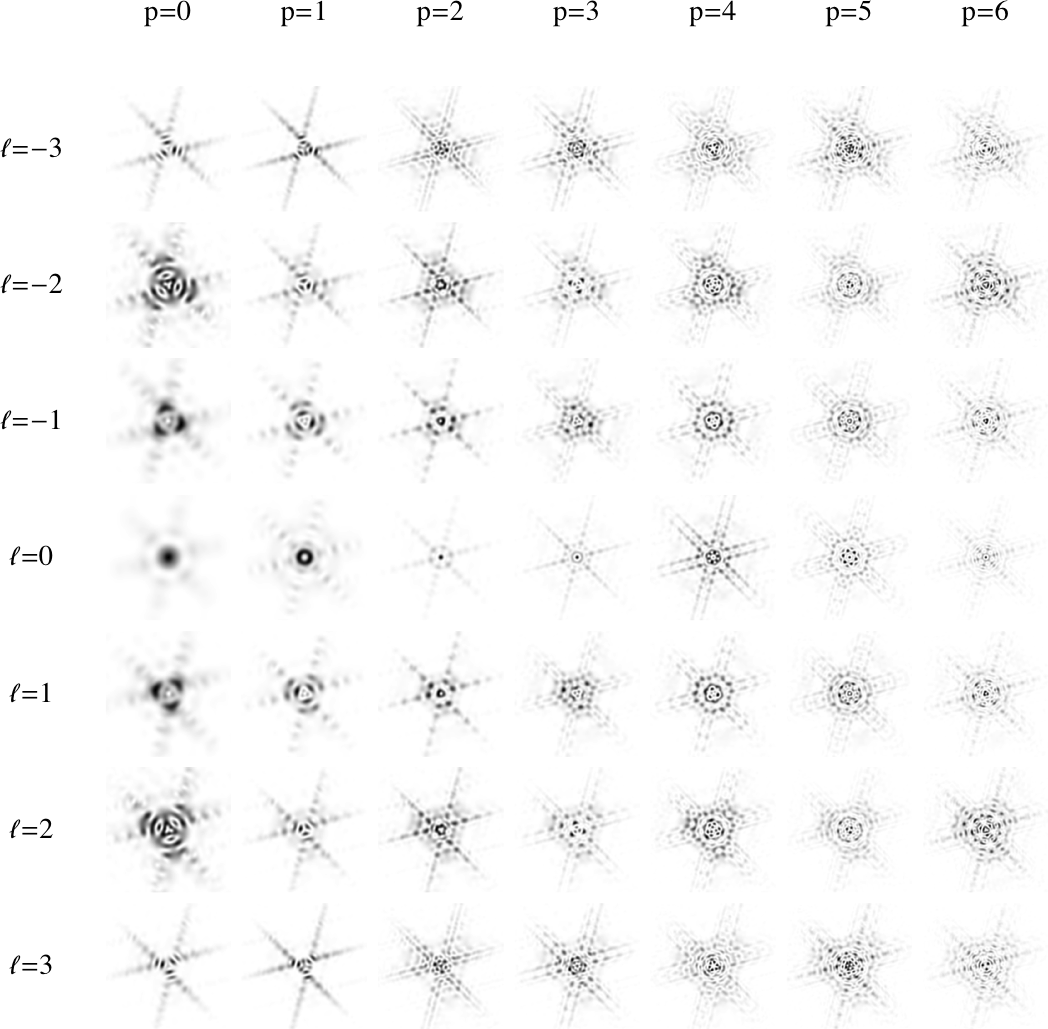} 
   \caption{Far-field intensity diffraction  patterns of Laguerre-Gaussian beams with different radial $p$ and azimuthal $\ell$ indices after diffraction from the triple triangular slit aperture (Fig. \ref{fig:L1P1}b). For clarity  higher intensity is represented by darker shade.}
   \label{fig:LP}
\end{figure}

In Fig.~\ref{fig:LP} we can observe the diffraction patterns of different beams. In general, these patterns show a complex interference interaction between the waist of the beam, the size and width of the triangular slit apertures, the number of triangular slit apertures and the azimuthal and radial indices of the Laguerre-Gaussian beams. 

\subsection*{S2 Optical eigenmode decomposition}

The optical eigenmode decomposition is based on the definition of the total intensity measured in a given region of interest (ROI).  Here we consider a superposition  of LG beams described by ${ E}=\sum_{j=1}^{N_1} a_j  {E}_j$ with ${ E}_j=u_2(\ell_j,p_j)$.
The total intensity, $I$, incident in the region of interest  is given by
\begin{equation}
I=\int_{ROI}  \;{ E}\cdot{ E}^*\;dxdy =\sum_{j,k} a^*_kM^{(0)}_{kj}a_j
\label{ROI}
\end{equation}
where $M^{(0)}_{kj}=\int_{ROI}  \;{ E_j}\cdot{ E_j}^*\;dxdy$.
Matrix $M^{(0)}_{kj}$ can be decomposed into a set of eigenvectors delivering an orthonormal set of optical eigenmodes defined as:
\begin{equation}
 {\mathbb{E}_q} =\frac{1}{\sqrt{\lambda_q}} \sum_j v_{q j}{ E}_j
  \label{OE2}
\end{equation}
with $ \sum_j M_{jk}{ v}_{q j}=\lambda_q { v}_{q k}$ and where $\lambda_q$ is an eigenvalue  and ${v}_{q j}$ the associated eigenvector.

\subsection*{S3 Wavefront correction using optical eigenmodes}

\begin{figure}[htbp] 
   \centering
   \includegraphics[scale=.3]{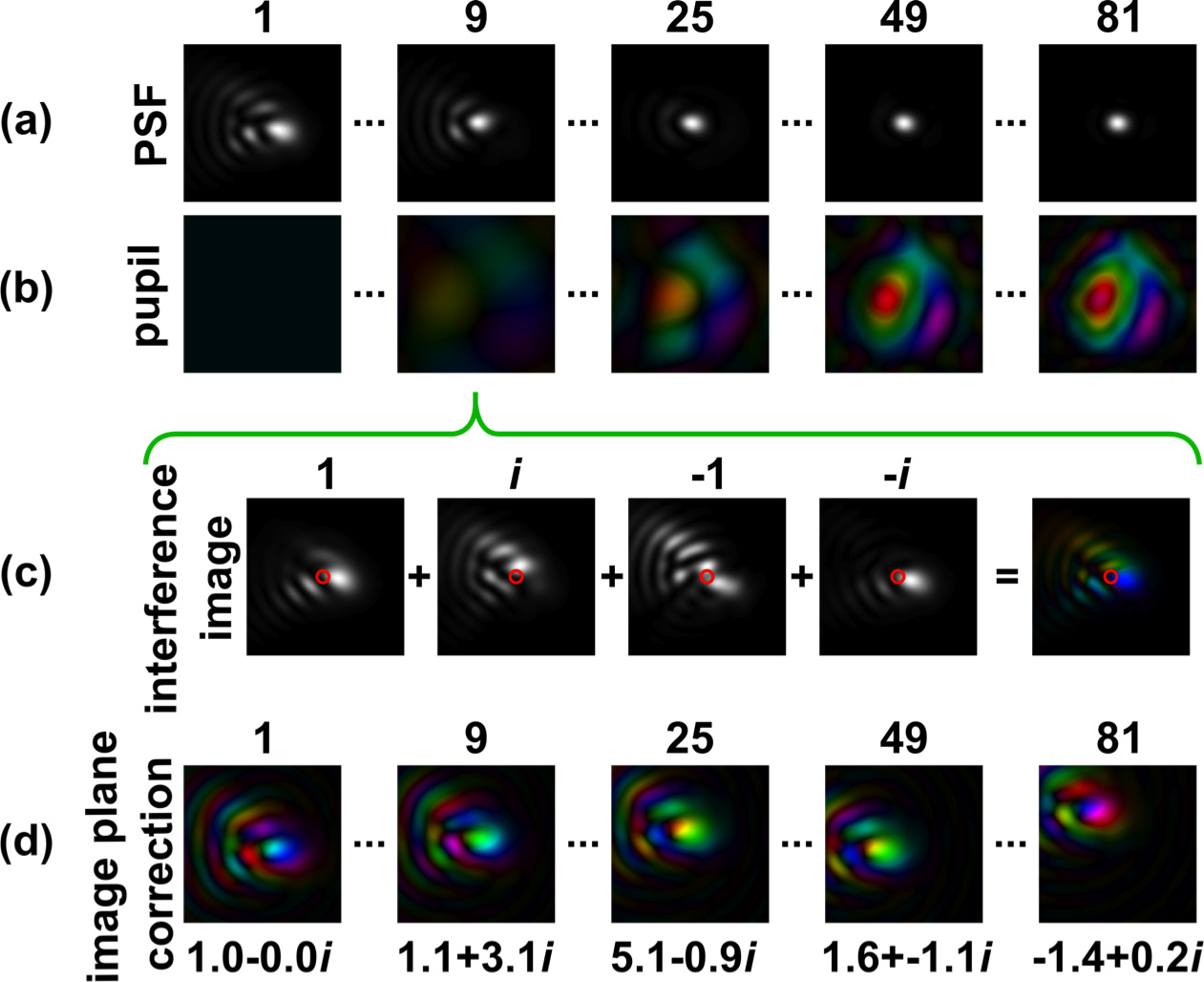}
   \caption{Description of the probing process of the coefficients of the intensity operator matrix. (a) The aberration corrected spot size shown after steps 1, 9, 25, 49, and 81 steps of the algorithm. (b) The corresponding estimate of the pupil aberration; brightness indicates amplitude and the hue of the colour indicates phase. (c) Detail of the calculation of step 9. Four intensity images are taken for which the probe is shifted in steps of $\pi/2$ with respect to the reference, and permit the determination of the complex field at the image plane. (d) The corrective adjustment of the image plane field at the five representative steps of the algorithm.}
   \label{fig:aberrationMeasurement}
\end{figure}

To mitigate the effect of optical aberrations inherent in any optical set-up, we employ an eigenmode wavefront correction algorithm based on equations~(\ref{ROI}-\ref{OE2}); however,
adapted for an experimental setting. We measure the intensity operator $M_{jk}$ describing the linear transfer between the SLM plane and the imaging plane by interfering a reference beam (${\rm E}_{\rm ref}$) with a set of probe beams ${\rm E}_{j}$. The progress of the algorithm is exemplified graphically in Fig.~\ref{fig:aberrationMeasurement}a and b for the representative probe beams 1, 9, 25, 49, and 81. Note that an accurate superposition of the reference and a probe beam can be obtained efficiently by designating half of the SLM pixels to each beam~\cite{Spalding2008fk}. As depicted in Figure~\ref{fig:aberrationMeasurement}c, the phase between the two beams is varied in four steps of $\pi/2$ and the complex field is retrieved using the equations:
\begin{eqnarray}
{\rm F}_{j}=\frac{1}{4} \sum^{3}_{p=0} e^{i2 \pi p/4} f({\rm E}_{\rm ref}+ e^{-i2\pi p/4}{\rm E}_{j})
\label{matrix2}
\end{eqnarray}
where the term $f({\rm E})$ corresponds to intensity measured at the detector in the presence of field $\rm E$. Figure~\ref{fig:aberrationMeasurement}c shows the four intensity images used in the calculation of probe step 9, together with the complex image plane field calculated from the four intensity images using equation~(\ref{OE2}). The procedure can be seen as a phase sensitive lock-in technique where the reference beam ${\rm E}_{\rm ref}$ corresponds to a reference signal with respect to which the phase and amplitude of ${\rm E}_j$ is measured. Using these measures we can define the intensity operator $M^{(0)}_{jk}=F^*_kF_j$. Its principal eigenvector defines the optical eigenmode that delivers the largest intensity on the detector whilst maintaining the total power in the system. The phase only part of the SLM mask implementing this eigenmode delivers the wavefront correction mask~(see \cite{Mazilu:2011uf} for a discussion on the number of probes used).

The top row of images in Figure.~\ref{fig:aberrationMeasurement} shows the spot after probing steps 1, 9, 25, 49, and 81 respectively. The false colour images on the second row show the complex correction applied by the SLM. As before, brightness indicates amplitude and hue indicates phase. The complex correction is calculated as a linear combination of the probe beams with coefficients determined by interfering the probe beam with a reference beam. This is exemplified for step 9 in Figure~\ref{fig:aberrationMeasurement}(c,d). The probe beams, linear phase gradients in this case, are phase shifted in steps of $\pi/2$ and interfered with the reference beam. For clarity, in this example the unmodulated beam is chosen as the reference. The four phase modulated probe beams are shown in colour coding in the top half of the second part of Figure~\ref{fig:aberrationMeasurement}. The four interference images are displayed below the corresponding probe patterns. We aim to maximise the intensity on the central image pixel by introducing a corrective pupil modulation that is a linear combination of the probe patterns. In this step of the algorithm we must therefore determine the coefficient of probe beam 9. This coefficient can be determined via \emph{lock-in amplification} from the intensity values measured at that pixel for the four phase shifts. The lock-in amplification determines the amplitude and phase of the complex coefficient. Since the intensity operator matrix is of rank 1, the largest eigenvector is simply the complex conjugate of this coefficient.


\begin{figure}[htbp] 
   \centering
   \includegraphics[scale=.33]{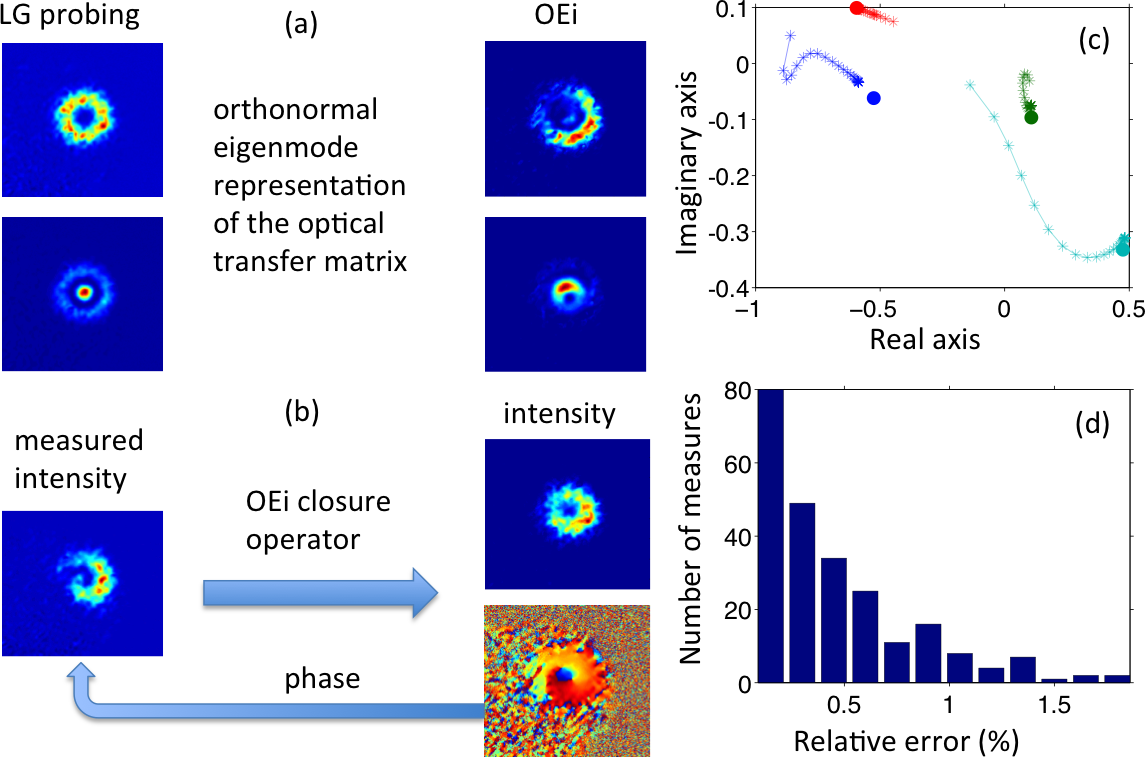}
   \caption{(a) Example LG input beam probing the optical system and resulting optical eigenmodes.  (b) Schema representing the generalised Gerchberg-Saxton (GS) algorithm based on the closure relation defined by the optical eigenmodes from part (a). (c) Convergence plot of the GS algorithm in the complex plane representing the amplitude and phases of the LG beam superposition ($\ell=[0,1]$ and $p=[0,1]$). (d) Relative detection error for a set of 100 random superpositions. We observe that 80\% of tests are below the 2\% relative error mark defined as $|c_i-m_i|^2$ where $c_i$ and $m_i$ are the, encoded respectively measured, complex superposition coefficients.}
   \label{fig:GS}
\end{figure}

\subsection*{S4 LG beam superposition phase retrieval using optical eigenmode Gerchberg-Saxton algorithm}

The PCA method employed in this letter is able to retrieve the real amplitude of a superposition of LG beams. To deduce the relative phase between the different LG beam constituents  we use a method based upon a generalised Gerchberg-Saxton~\cite{GS} algorithm (GS) working in tandem with an optical eigenmode decomposition~\cite{DeLuca:2011jl,Mazilu:2011uf}.
These eigenmodes correspond to a family of electromagnetic fields that in the imaging plane form, by construction, an orthogonal set of fields (see supplement S2 for details).  The OEi can be measured experimentally by probing the optical system with a set of interfering beams created with the SLM~\cite{DeLuca:2011jl} (schematically illustrated in figure~\ref{fig:GS}a).  If this set forms a large enough set of independent beams then the deduced OEi can be considered to form a complete basis accounting for all the optical degrees of freedom of the system. If the probing set is not large enough then the closure relationship of the OEi base corresponds to a projection operator, account only for a restricted set of degrees of freedom of the system. Here, we use the LG beams considered for the superposition to measure this closure operator that subsequently can be used to describe the propagation of the electromagnetic field as a superposition of the same considered LG beams. Replacing the Fourier transform in the GS algorithm with this closure operator delivers a constrained iterative GS procedure that can be used to deduce the complex superposition coefficients of the incident LG beams~\ref{fig:GS}b. Figure~\ref{fig:GS}c,d shows that this iterative process converges to the correct relative phases in more than 60\% of cases. This convergence is helped through the use of the amplitudes measured using the PCA method as starting points.

\end{document}